\documentclass[twocolumn,showpacs,amsfonts]{revtex4-1}

\usepackage{amsmath}
\usepackage{latexsym}
\usepackage{mathrsfs}
\usepackage{float}
\usepackage{amssymb}
\usepackage{amsfonts}
\usepackage{graphicx}
\usepackage{textcomp}
\usepackage{hyperref}
\usepackage{mathrsfs}
\usepackage{pifont}
\usepackage{color}
\usepackage{latexsym}

 1

\newcommand{\dummy}

\begin{document}
\title{Finite-size Scaling Study of Shear Viscosity Anomaly at Liquid-Liquid Criticality}
\author{Sutapa Roy$^{1,2}$ and Subir K. Das$^{*}$}
\affiliation{$1$ Theoretical Sciences Unit, Jawaharlal Nehru 
Centre for Advanced Scientific Research, Jakkur P.O, 
Bangalore 560064, India\\
$2$ Max-Planck-Institut f\"ur Intelligente Systeme, Heisenbergstr. 3, 70569 Stuttgart, 
Germany and Institut f\"ur Theoretische Physik IV, Universit\"at Stuttgart, 
Pfaffenwaldring 57, 70569 Stuttgart, Germany}
\date{\today}

\begin{abstract}
We study equilibrium dynamics of a symmetrical binary Lennard-Jones fluid mixture 
near its consolute criticality. Molecular dynamics simulation results for shear 
viscosity, $\eta$, from microcanonical ensemble are compared with those from canonical 
ensemble with various thermostats. It is observed that Nos\'{e}-Hoover thermostat 
is a good candidate for this purpose and so, is adopted for the quantification of 
critical singularity of $\eta$, to avoid temperature fluctuation (or even drift) 
that is often encountered in microcanonical simulations. Via finite-size scaling 
analysis of our simulation data, thus obtained, we have been able to quantify even 
the weakest anomaly, of all transport properties, that shear viscosity exhibits and 
confirm the corresponding theoretical prediction.
\end{abstract}
\maketitle

\section{\label{intro}Introduction}
\par
\hspace{0.2cm}
Knowledge of the behaviors of equilibrium transport properties, with the variation of 
various thermodynamic parameters, is crucial even to the understanding of 
nonequilibrium dynamics \cite{onuki1,bray}. In computer simulations, reliable 
calculations of collective transport properties \cite{onuki1,hansen}, e.g., shear 
($\eta$) and bulk ($\zeta$) viscosities, thermal diffusivity ($D_T$), mutual 
diffusivity ($D_{AB}$) in a ($A+B$) binary mixture, etc., are extremely difficult 
due to lack of self averaging \cite{onuki1,hansen,allen}. Simulation studies of 
dynamic critical phenomena, because of this reason, have started only recently 
\cite{yeth1,yeth2,chen,das2006,dasjcp2006,roy1,roy2}.

\par
\hspace{0.2cm}
In the vicinity of a critical point, various static and dynamic properties exhibit 
power-law singularities \cite{onuki1,zinn,hohenberg}. E.g., in static critical 
phenomena, the correlation length ($\xi$), order parameter ($m$) and susceptibility 
($\chi$) behave as \cite{onuki1,zinn} 
\begin{equation}\label{statexp}
\xi \approx \xi_0^{\pm} \epsilon ^{-\nu},~m\approx \hat{B} \epsilon ^{\beta}, 
~\chi\approx \Gamma^{\pm}\epsilon ^{-\gamma};~
\epsilon=\frac{|T-T_c|}{T_c},
\end{equation}
where $T$ is the temperature, $T_c$ being its value at the critical point. The 
superscripts $\pm$ on the amplitudes $\xi_0$ and $\Gamma$ signify singularities 
irrespective of which side of $T_c$ one approaches it. In dynamics 
\cite{onuki1,hohenberg,kadanoff,anisimov,onuki2,olchowy,folk,hao,jkb1,jkb2,sengers,
burstyn1,burstyn2,landau}, the relaxation time ($\tau$) and the above mentioned 
transport quantities have the critical behaviors
\begin{equation}\label{dynexp}
\tau\sim\xi^{z},~\eta\sim\xi^{x_{\eta}},~\zeta\sim\xi^{x_{\zeta}},~D_{T,AB}\sim\xi^{-x_D}.
\end{equation}

\par
\hspace{0.2cm}
The values of the critical exponents do not depend upon the atomistic details 
of the systems. In case of static properties, these are even independent of type 
of transitions, i.e., demixing transitions in binary solids and fluids, 
vapor-liquid transitions or para- to ferromagnetic transitions will have same 
values of the exponents, if the system dimensionality and number of order-parameter 
components are same, alongside inter-particle interactions being of similar range. For 
short range interactions, with one component order parameter, the universality 
belongs to the Ising class for which the values of the exponents in space 
dimension $d=3$ are \cite{zinn}
\begin{eqnarray}\label {static2}
\nu=0.63,~\beta=0.325,~\gamma=1.239.
\end{eqnarray}

\par
\hspace{0.2cm}
Such universalities exist in dynamics as well, though less robust. E.g., 
vapor-liquid and liquid-liquid transitions are expected to belong to the same 
universality \cite{hohenberg,kadanoff,anisimov,onuki2,olchowy,folk,hao,jkb1,jkb2,
sengers,burstyn1,burstyn2,landau}, with $z\simeq 3.068$, which is different 
from that of a para- to ferromagnetic transition \cite{landau}, having 
$z\simeq 2.15$. This difference is due to the nonconservation of order parameter
in the ferromagnetic case. In addition to standard finite-size effects, encountered in static 
critical phenomena, the high value of $z$, particularly for fluid criticality, 
makes the study of dynamic critical phenomena significantly more difficult. 
This phenomenon, referred to as the critical slowing down, can be appreciated from 
the fact that the relaxation time at criticality, where $\xi$ scales with the 
system size $L$, diverges as \cite{landau}
\begin{equation}\label{relax}
\tau \sim L^z.
\end{equation}

\par
\hspace{0.2cm}
The values of the other dynamic exponents in $d=3$ fluid universality class are 
\cite{onuki1,hao,jkb1,jkb2,olchowy}
\begin{equation}\label{dyn2}
x_{\eta}\simeq 0.068,~x_{\zeta}\simeq 2.89,~x_D\simeq 1.068,
\end{equation}
for $x_{\zeta}$ the value being slightly higher for a liquid-liquid transition. 
For the latter, singularities of $D_{AB}$ and $\zeta$ were 
recently verified via molecular dynamics (MD) simulations 
\cite{das2006,dasjcp2006,
roy1,roy2}. 
However, quality of data were not appropriate enough to draw a conclusion 
on the anomaly of $\eta$, because of the tiny value of $x_{\eta}$.

\begin{figure}
\centering
\includegraphics*[width=0.31\textwidth,height=0.3\textwidth]{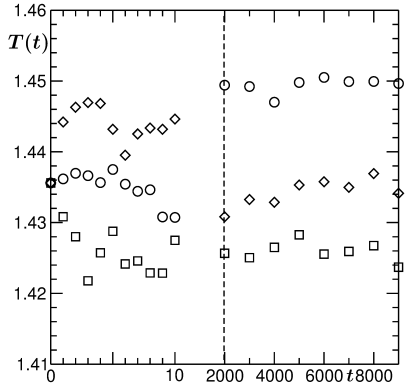}
\caption{\label{fig1}
Plots of temperature as a function of time, during microcanonical molecular 
dynamics runs of the binary Lennard-Jones mixture. The system size is $L=14$. 
Before each of these runs, the system was appropriately equilibrated for 
long enough time via Monte Carlo runs in canonical ensemble and further, 
thermalized via molecular dynamics runs with Andersen thermostat. The initial 
temperature was set to a value reasonably close to the critical one.}
\end{figure}

\par
\hspace{0.2cm}
Fluid transport properties, via MD simulations \cite{allen,frenkel,rapaport}, 
are traditionally calculated in microcanonical ensemble (constant NVE, $N$ 
being the total number of particles, $V$ the volume and $E$ energy) where 
hydrodynamics is ideally satisfied. However, temperature in this ensemble 
is not perfectly controlled which is nondesirable for the quantification of 
critical singularity, particularly when expected anomaly is weak. The 
fluctuation of temperature in NVE ensemble, during MD simulations of 
our binary fluid model (to be defined soon), is shown in Fig.~\ref{fig1}. 
There we have plotted $T$ as a function of time ($t$) for a few different 
runs. It is common experience, as seen here, that in long time limit, 
the value of $T$ settles down to a nondesirable number. 
For more realistic models (than the one we use here), the temperature
drift can be stronger than seen in Fig.~\ref{fig1}.

\par
\hspace{0.2cm}
To avoid the above mentioned problem, we have planned to calculate 
$\eta$ in canonical ensemble as well, with hydrodynamics preserving thermostat. 
In addition to accurate computation, an objective of the work is to demonstrate 
that critical dynamics in fluids can be studied in canonical ensemble.
The suitable candidates for this purpose are Nos\'{e}-Hoover thermostat 
(NHT) \cite{frenkel}, dissipative particle dynamics \cite{stoya,niku}, 
multiparticle collision dynamics \cite{gomp}, etc. For technical reasons, 
related to temperature control, we will adopt the NHT for this work.
Our simulation data from NHT, when analyzed via finite-size scaling
(FSS) \cite{landau,fisher_fss}, show excellent agreement 
with the theoretically predicted singularity, quoted in Eq. (\ref{dyn2}). 
As mentioned earlier, this is the first such confirmation of the critical 
anomaly of shear viscosity.

\par
\hspace{0.2cm}
The rest of the paper is organized as follows. In Section II we discuss
the model and methods. Results are presented in Section III. Finally,
Section IV concludes the paper with a brief summary.

\section{Model and Methods}\label{model}
\par
\hspace{0.2cm}
We consider a binary fluid model 
\cite{das2006,dasjcp2006,roy1,roy2} where particles, all of same mass $m$, 
at continuum positions ${\vec r}_i$ and ${\vec r}_j$, interact via the 
Lennard-Jones (LJ) pair potential 
\begin {eqnarray}\label{LJ1}
u(r=|{\vec r}_i-{\vec r}_j|)=4\varepsilon_{_{\alpha\beta}}
\Big[\Big(\frac{d_0}{r}\Big)^{12}-\Big(\frac{d_0}{r}\Big)^{6}\Big],
\end{eqnarray}
where $d_0$ is the particle diameter, same for all, and 
$\varepsilon_{\alpha\beta}~[\alpha,\beta=A,B]$ are the interaction 
strengths. We set $\varepsilon_{_{AA}}=\varepsilon_{_{BB}}=
2\varepsilon_{_{AB}}=\varepsilon$ which, in addition to facilitating 
phase separation, makes the model perfectly symmetric. For computational 
benefit, we truncate the potential at an inter-particle distance 
$r=r_c=2.5d_0$ and work with the shifted and force-corrected form 
\cite{allen}
\begin {eqnarray}\label{LJ2}
U(r)=u(r)-u(r_c)-(r-r_{_c}){\frac {du}{dr}}\Big|_{{r}=r_{_c}}.
\end{eqnarray}
The number density of particles $\rho$ ($=N/L^3$) was fixed to $1$, 
a value high enough to avoid any overlap with a vapor-liquid transition.

\begin{figure}
\centering
\includegraphics*[width=0.31\textwidth,height=0.3\textwidth]{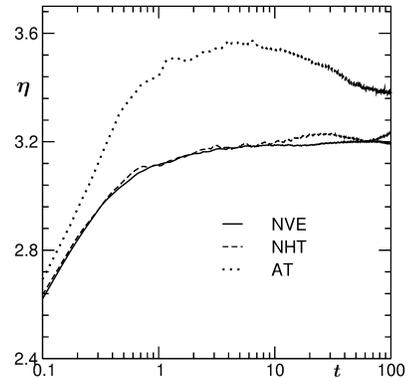}
\caption{\label{fig2}
Plot of $\eta (t)$ as a function of $t$ at $T=3.25$. Results from 
microcanonical MD and canonical MD with NHT and AT are shown. The value of
$\eta$ can be obtained from the plataea of these plots.}
\end{figure}

\par
\hspace{0.2cm}
Static quantities for this model were studied 
\cite{das2006,dasjcp2006,roy1,roy2} via Monte Carlo \cite{landau} 
(MC) simulations in semi-grand canonical ensemble \cite{landau,frenkel} 
(SGMC) where, in addition to standard displacement moves, identity 
switches $A \rightarrow B \rightarrow A$  are also tried. This allows 
one to record the fluctuations of concentration 
$x_\alpha$ ($=N_\alpha/N$, $N_\alpha$ being the number of particles 
of species $\alpha$) of either species and thus distribution 
$P(x_{\alpha})$. As we will discuss later, this enables estimation of 
phase diagram, in addition to several other static quantities.

\par
\hspace{0.2cm}
Transport quantities were calculated using the MD simulations in 
NVE as well as in Canonical ensembles. For the latter we have used 
NHT as well as Andersen thermostat (AT) \cite{frenkel}. In AT, to keep 
the temperature constant, randomly chosen particles are made to 
collide with a fictitious heat bath, i.e, they are assigned new velocities 
in accordance with the desired temperature. Thus hydrodynamics is not expected 
to be preserved by AT. For NHT, we have solved the standard 
deterministic equations of motion \cite{frenkel}
\begin{eqnarray}\label {NHT1}
m_i\dot {r_i} &=& p_i,\\
\dot {p_i} &=& -\frac{\delta U_i}{\delta r_i}-\Xi p_i, \\
\dot \Xi &=& \Big(\sum_{i=1}^N p_i^2/m_i-3N{k_B} T\Big)/Q,
\end{eqnarray}
where $k_B$ is the Boltzmann constant, $\Xi$ is a time dependent drag 
and $Q$ is a constant providing the strength of coupling of the system with the 
thermostat. The value of $Q$ was set to $1$ and also for $\Xi$ we have 
started with unity in each MD run. In both the ensembles, Verlet velocity 
algorithm was used to integrate the equations of motion, with the time 
step of integration being set to $0.005t_0$, where 
$t_0$ ($=\sqrt{md_0^2/\varepsilon}$) is the LJ time unit.

\par
\hspace{0.2cm}
We computed shear viscosity using the Green-Kubo (GK) and Einstein relations \cite{hansen}. 
The GK relation is given by 
\begin{eqnarray}\label{shear1}
\eta(t)={\left(\frac {t_{_0}^3 \varepsilon}{Vd_0Tm^2}\right)
\int_0^t dt' 
\langle {\sigma_{xy}}(t'){\sigma_{xy}}(0)\rangle},~~~
\end{eqnarray}
where $\sigma_{xy}$ are the off-diagonal elements of the pressure tensor, computed as
\begin{eqnarray}\label{shear2}
\sigma_{xy}(t)=\sum_{i=1}^N\Big[mv_{ix}v_{jy}+
\frac{1}{2}\sum_{j(\ne i)}(x_i-x_j)F_{jy}\Big],~~~
\end{eqnarray}
with $v_{_{ix}}$ being the $x$ component of the velocity of $i$th particle 
and $F_{jy}$ the $y$ component of force acting 
on the $j$th particle due to all others. The Einstein formula for 
the calculation of $\eta$ is 
\begin{eqnarray}\label{shearmsd1}
\eta(t)=\left(\frac {t_{_0}^3\varepsilon}{2k_{_B}tVd_0T m^2}\right)
\langle |Q_{_{xy}}(t)-Q_{_{xy}}(0)|^2\rangle,
\end{eqnarray}
where the generalized displacement $Q_{xy}$ has the form 
\cite{hansen}
\begin{eqnarray}\label{shearmsd2}
Q_{_{xy}}(t)=\sum_{i=1}^N {x_{_i}(t)v_{iy}(t)}.
\end{eqnarray}
Both Eq. (\ref{shear1}) and Eq. (\ref{shearmsd1}) provide similar results, 
but in this paper we present the ones obtained only from the Einstein relation. 

\begin{figure}
\centering
\includegraphics*[width=0.31\textwidth, height=0.3\textwidth]{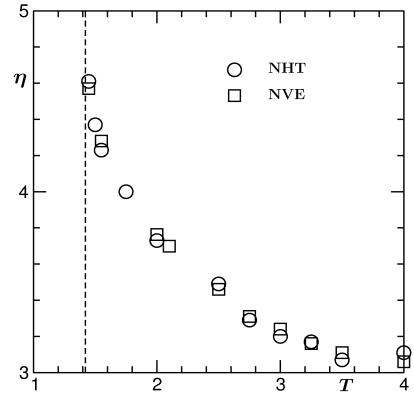}
\caption{\label{fig3}
The shear viscosity is plotted as a function of $T$. Results from 
NVE and NHT calculations are shown. The system size used was $L=10$. 
The dashed line stands for $T_c=1.421$.}
\end{figure}

\par
\hspace{0.2cm}
All simulations were performed in cubic boxes of 
volume $V$($=L^3$, $L$ being in units of $d_0$), with periodic boundary 
condition in each direction. Results are presented after averaging 
over multiple initial configurations. For phase behavior, at each 
temperature, $6$ different initial configurations were averaged over, 
while this number was from $400$ to $650$, depending on the system size, 
for shear viscosity. For 
dynamics we have stuck to the critical composition $x_{\alpha}^c=1/2$, 
the value being set by the symmetry of the model. For the rest of our 
paper, the values of $m$, $\varepsilon$, $d_0$, $t_0$ and $k_B$ were 
fixed to unity.

\section{Results}\label{results}
\par
\hspace{0.2cm}
In Fig.~\ref{fig2} we plot $\eta (t)$ as a function of $t$, obtained 
via Eq. (\ref{shearmsd1}), at $T=3.25$. Results are presented from NVE and 
NVT ensembles. For the latter, calculations using both NHT and AT are 
included. It is clearly seen that NHT produces result comparable with 
NVE ensemble. As discussed, AT does not meet the momentum conservation
condition. Deviation of AT result from the others provide further
confidence that the matching of NHT and NVE is not accidental.
Final values of $\eta$ were obtained from the flat regions 
in $\eta(t)$ vs $t$ plots. 

\par
\hspace{0.2cm}
In Fig.~\ref{fig3} we show the variation of $\eta$ as a function of $T$, for $L=10$. 
Critical enhancement is visible.
Results from NVE and NHT are shown which agree nicely with each other. 
Being encouraged by the good agreement 
between them, we adopt the latter for quantification of critical 
singularity via FSS. At this stage we do not attempt to quantify the
critical singularity using these data, fearing
finite-size effects. We introduce the FSS method below, via discussion of 
the phase behavior.

\begin{figure}
\centering
\includegraphics*[width=0.32\textwidth, height=0.3\textwidth]{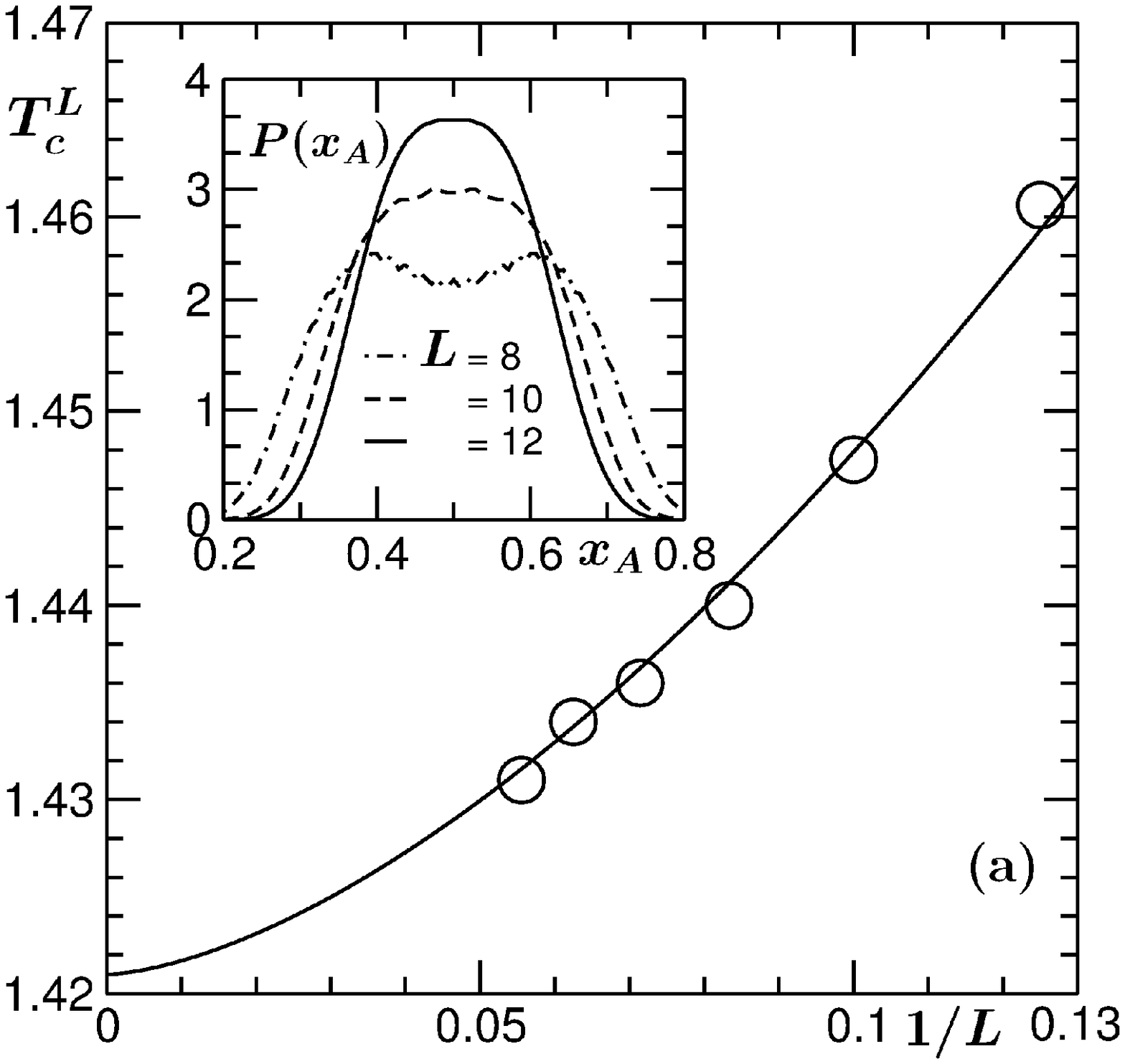}
\vskip 0.5cm
\includegraphics*[width=0.315\textwidth, height=0.3\textwidth]{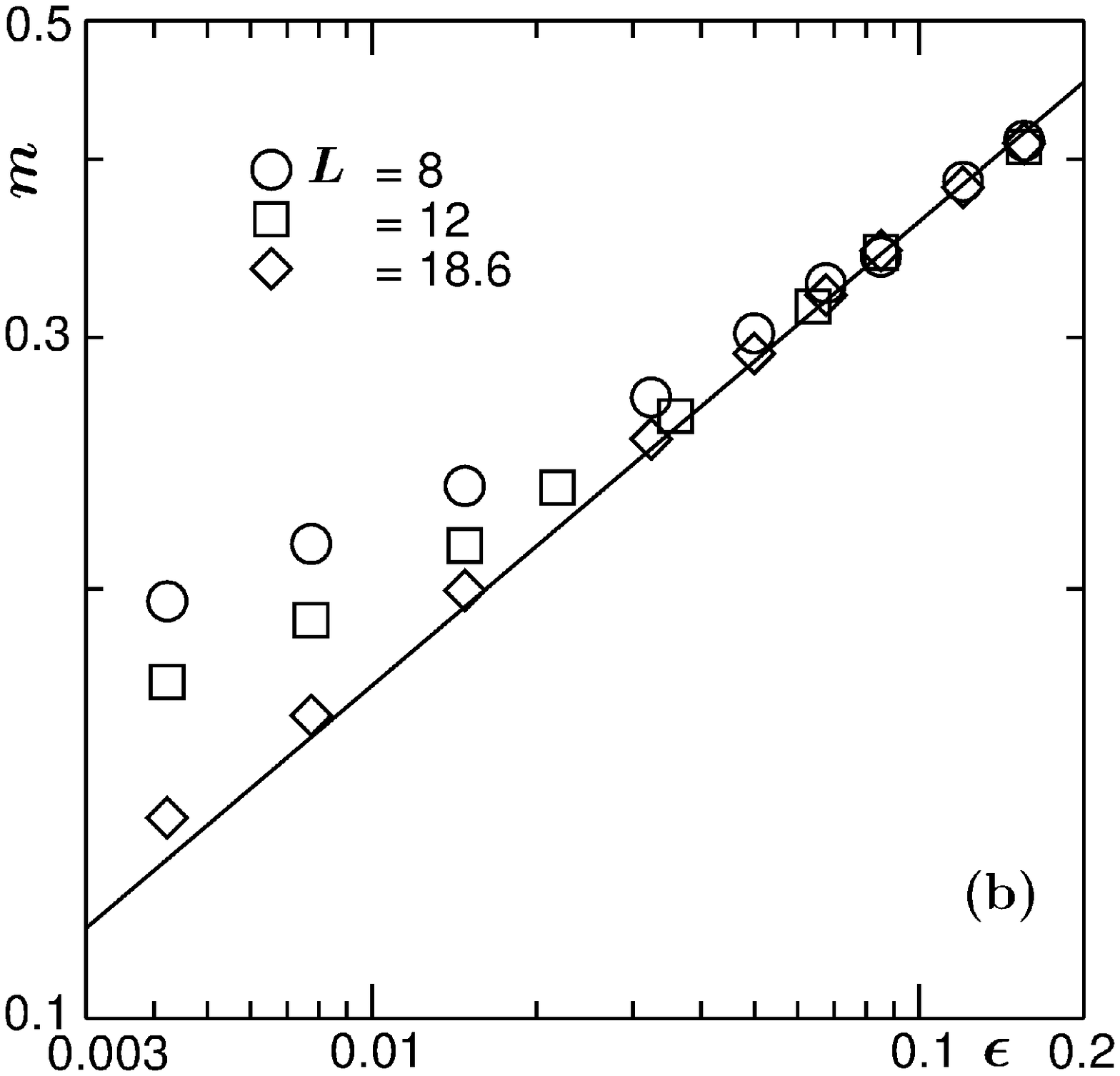}
\caption{\label{fig4}
(a) Plot of $T_c^L$ as a function of $1/L$. The continuous line there is a 
fit to the scaling form (\ref{scltcl}). This line meets the ordinate at 
$T_c=1.421$ for $L=\infty$. Inset: Probability distributions $P(x_A)$ 
is plotted vs $x_A$, for a few different system sizes at $T=1.4475$. 
(b) Order parameter $m$, defined in Eq. (\ref{opdef}), is plotted vs 
$\epsilon$, on log-log scales, for a few different system sizes, as 
indicated. The continuous line has Ising critical behavior.}
\end{figure}

\par
\hspace{0.2cm}
Like all other quantities, phase diagram also suffers from finite-size 
effects, in the close vicinity of a critical point. For a binary mixture, 
the demixing phase diagram can be obtained from the SGMC simulations in 
the following way. The probability distribution $P(x_{\alpha})$ will 
have a double peak structure below the critical point. The locations of 
these peaks will provide points for the coexistence curve. Thus, these 
peaks should merge at the critical point. However, the merging temperatures 
will be different for different values of $L$, providing finite-size 
critical points, $T_c^L$, with $T_c=T_c^{L=\infty}$. A plot of $T_c^L$, 
as a function of inverse system size, is shown in Fig.~\ref{fig4}(a). In 
the inset of this figure, we have shown a few distribution functions, 
for concentrations of $A$ particles, at the same temperature but for 
different system sizes. There it is seen, even though at this temperature 
there is phase coexistence for $L=8$, a system with $L=12$ is certainly 
in the homogeneously mixed state. This chosen temperature corresponds to 
$T_c^L$ for $L=10$.

\par
\hspace{0.2cm}
Since $\xi$ scales with $L$ at criticality, the finite-size critical 
behavior of $T_c^L$ is given by \cite{landau,fisher_fss}
\begin{equation}\label{scltcl}
T_c^L-T_c \sim L^{-1/\nu}.
\end{equation}
The continuous line in Fig.~\ref{fig4}(a) is a fit of the simulation data 
set to the form in Eq. (\ref{scltcl}). Because of the expectation that a 
LJ system should belong to the Ising universality class, we have fixed 
$\nu$ to $0.63$. This provides $T_c=1.421$ which is in good agreement with 
estimates via other methods \cite{das2006,dasjcp2006}.

\begin{figure}
\centering
\includegraphics*[width=0.3\textwidth]{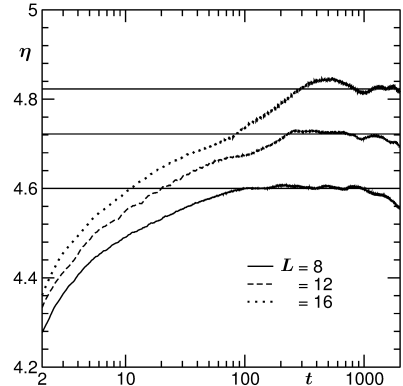}
\caption{\label{fig5}
Plot of $\eta(t)$ as a function of $t$ at $T=T_c^L$ of three different 
system sizes. Estimation of $\eta$ from the flat regions, for each of 
these systems, is demonstrated.}
\end{figure}

\par
\hspace{0.2cm}
Again, because of scaling of $\xi$ with $L$ at criticality, when 
calculated at $T_c^L$, for various system sizes, we expect
\begin{equation}\label{shearfss}
\eta\sim L^{x_{\eta}}.
\end{equation}
Before getting into that exercise, we plot the order parameter $m$ as a 
function of $\epsilon$, for different system sizes, in Fig.~\ref{fig4}(b), 
on log-log scales. Here we have used $T_c=1.421$. The solid line in this 
figure corresponds to the critical singularity of $m$ with $\beta=0.325$, 
$m$ being defined as 
\begin{equation}\label{opdef}
m=|x_A^{\rm coex} -1/2|.
\end{equation}
Here $x_A^{\rm coex}$ is the concentration of $A$ particles either in 
$A$ or $B$-rich coexisting phase which, as already mentioned, can be 
obtained from the locations of the peaks in the inset of Fig.~\ref{fig4}(a), 
for $T < T_c^L$. As seen, with increasing system size, data become more 
and more consistent with the theoretical line. This provides further 
confidence on the estimations of $T_c^L$, presented in Fig.~\ref{fig4}(a).

\begin{figure}
\centering
\includegraphics*[width=0.315\textwidth,height=0.3\textwidth]{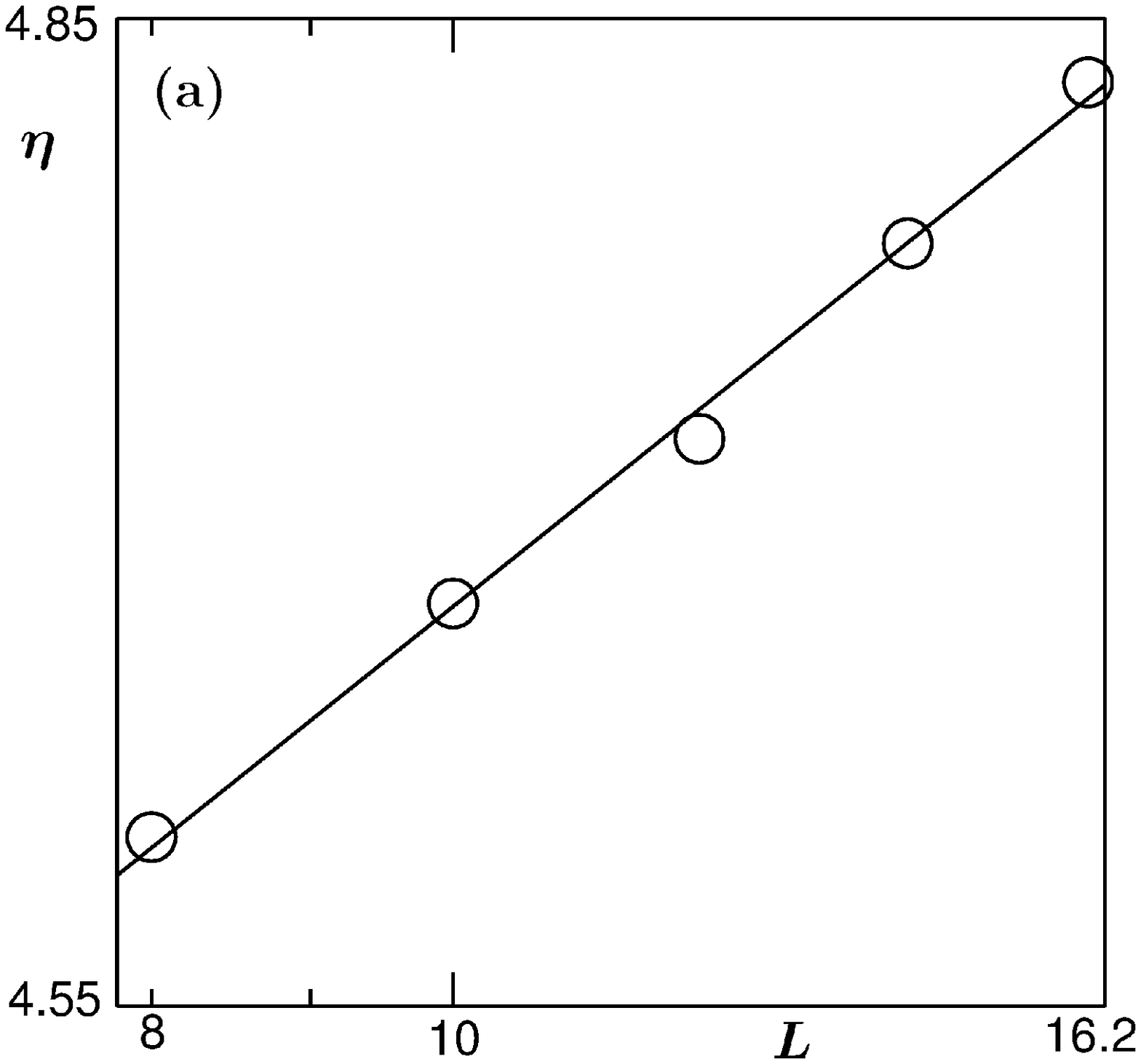}
\vskip 0.5cm
\includegraphics*[width=0.3\textwidth,height=0.3\textwidth]{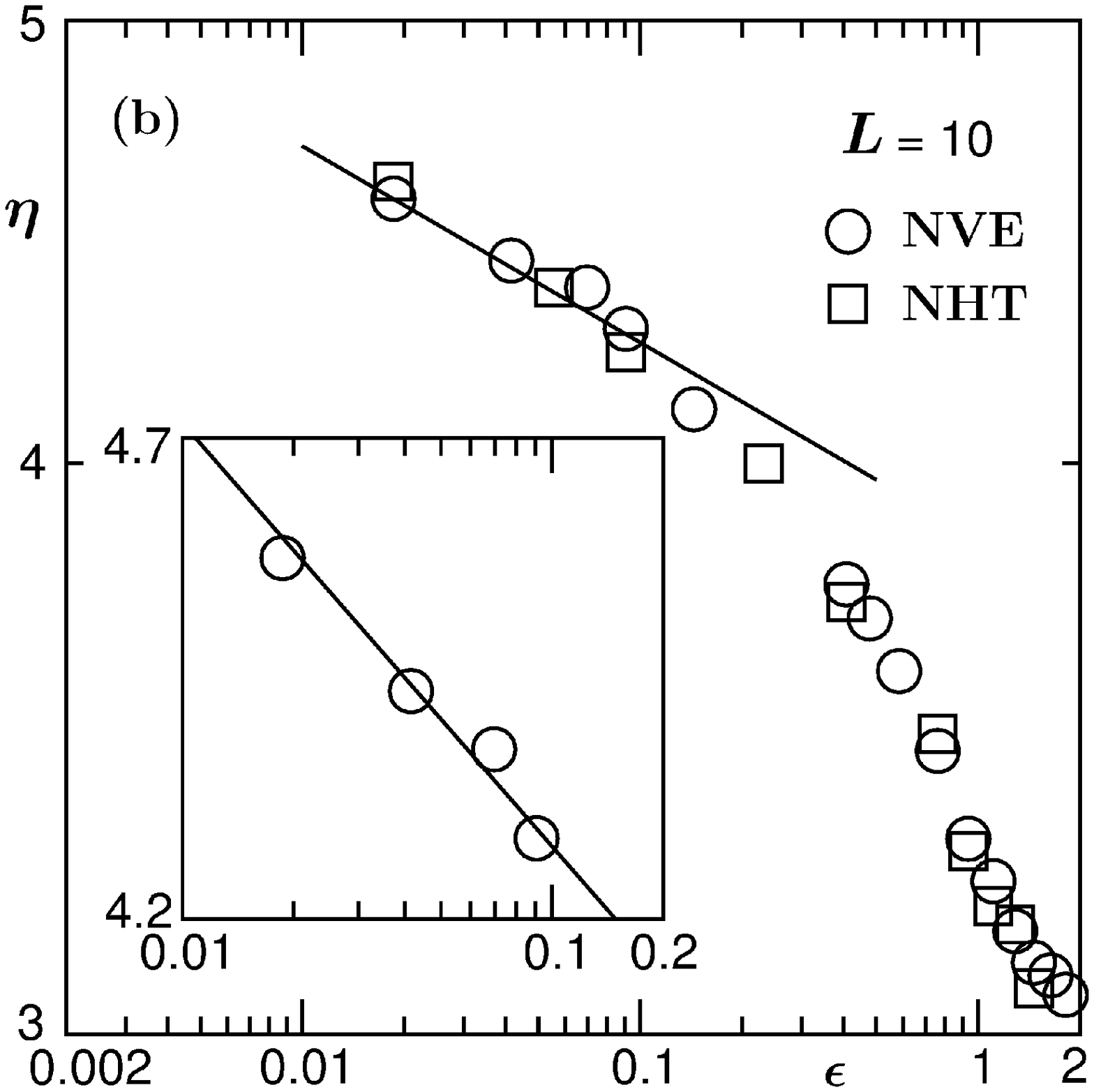}
\caption{\label{fig6}
(a) Finite-size scaling estimation of 
critical exponent $x_{\eta}$. Here we have plotted $\eta$ as a function 
of $L$. For each value of $L$, calculation was done at the corresponding 
$T_c^L$. A double-log scale is used. The continuous line there has the 
theoretical exponent $0.068$. (b) The NHT and NVE data from Fig.~\ref{fig3} are
plotted vs $\epsilon$, on double-log scale. The inset contains data
for only the four smallest values of $\epsilon$, from NVE calculations. 
The solid lines in (b)
correspond to the theoretical expectation.}
\end{figure}

\par
\hspace{0.2cm}
In Fig.~\ref{fig5}, we show $\eta (t)$ as a function of $t$, at $T_c^L$ 
of three different system sizes. The estimates of $\eta$ from the flat 
regions of these plots are demonstrated. These values are plotted in 
Fig.~\ref{fig6}(a), as a function of $L$. On the log-log scales, the data 
appear fairly linear, indicating a power-law behavior. The continuous 
line there is a fit to the form (\ref{shearfss}), providing $x_{\eta}\simeq 0.070$.
This is in excellent agreement with the value $x_{\eta}\simeq 0.068$, predicted
\cite{ferrell,jkbphysa,hao} by the theory of Ferrell and Bhattacharjee. 
For more reliability of a number from 
fitting, one, of course, needs data for more number of $L$ values. But 
due to lack of self averaging, computations of collective properties 
from MD simulations are extremely difficult task.

\par
\hspace{0.2cm}
Having confirmed the theoretical exponent, we move to check for the extent of
critical region. In Fig.~\ref{fig6}(b), we plot $\eta$, from both NVE and NHT 
calculations, as a function of $\epsilon$,
for $L=10$, on log-log scales. It appears that data only upto 10\% above $T_c$ are
consistent with the theoretical exponent, represented by the solid line. This, of
course, is in agreement with the experimental observations for both statics
and dynamics. For statics, of course, computer simulation results also
show enhancements only upto 10\% away from the critical temperature. 
Interestingly, this
is at deviation with the critical range of bulk viscosity and Onsager coefficient
for this particular model \cite{das2006,dasjcp2006,roy1,roy2}. For these quantities,
the expected power laws were observed upto even 100\% above $T_c$. In the inset of 
Fig.~\ref{fig6}(b) we show data only for four smallest values of $\epsilon$, from 
NVE calculation, which
show reasonably good consistency with the (solid) theoretical line having
exponent $x_{\eta}\nu=0.043$. A power law fit to this data set provides
$0.039$, i.e., $x_{\eta}\simeq 0.062$. This smaller value can well be due
to finite-size effects or statistical fluctuation. Note that the averaging statistics 
for these results were poorer than that in Fig. \ref{fig6}(a). 
In all these analysis, a background $\eta_b$ was not considered. Here note that, $\eta$ is
expected to exhibit a multiplicative critical anomaly, i.e., the enhancement of $\eta$ should be
proportional to $\eta_b$ \cite{newrev,dsf2007}. The current set of simulation results, however, are
inadequate to resolve this important issue which we leave out for future work.
The range of observation of critical enhancement, discussed above, in fact, 
depends upon whether one looks at
$\eta$ or $\eta/\eta_b$. The lack of knowledge about $\eta_b$ prevents us from the
latter exercise.

\section{Conclusion}\label{conclusion}
\par
\hspace{0.2cm}
In summary, we have studied the critical behavior of shear viscosity in 
a symmetrical binary Lennard-Jones fluid. Canonical ensemble 
calculations with Nos\'{e}-Hoover thermostat (NHT) provide shear viscosity 
comparable to the ones computed from microcanonical ensemble. 
Better momentum conserving thermostats are, of course, available \cite{schmid}.
But appropriate compromise between momentum conservation and temperature 
control is essential.

\par
\hspace{0.2cm}
Finite-size scaling analysis of our NHT simulation data show nice agreement 
with theoretical expectation. This is the first simulation confirmation of 
the critical anomaly of shear viscosity, irrespective of a liquid-liquid or 
vapor-liquid transition. Before this, to the best of our knowledge there 
were three simulation studies \cite{yeth2,dasjcp2006,roy2} on dynamic critical phenomena, 
where results for shear viscosity were also presented. In one of these 
studies \cite{dasjcp2006}, accepting the theoretical exponent, only the 
critical amplitude was estimated which had its relevance in obtaining a 
theoretical number for the critical amplitude of Onsager coefficient. 
In another study \cite{yeth2}, a statement about the upper bound of 
$x_{\eta}$ was made by using only two data points. These again are 
supportive of the difficulty one encounters in computational studies of 
such collective properties.

\par
\hspace{0.2cm}
Our conclusions apply to the asymptotic critical behavior of a liquid mixture \cite{newrev}. Here
note that the asymptotic range for viscosity for vapor-liquid transition is limited to
an extremely small range in the critical point vicinity and non-asymptotic corrections
become essential for the understanding of experimental results \cite{mold1999}. 
This is related to the fact,
as mentioned above, that the critical behavior of $\eta$ is proportional to $\eta_b$ which is
much larger in a liquid mixture than in a vapor-liquid transition.
\vskip 1.0cm

\section*{Acknowledgement}\label{acknowledgement} 
\hspace{0.2cm} SKD and SR acknowledge financial support from the Department of 
Science and Technology, India, via Grant No SR/S2/RJN-$13/2009$. 
SR is grateful to the Council of Scientific and Industrial 
Research, India, for their research fellowship.
\vskip 0.5cm
\par
$*$~das@jncasr.ac.in


\vskip 0.5cm
\begin{thebibliography}{100}

\bibitem{onuki1}A. Onuki, \textit{Phase
Transition Dynamics} (Cambridge University Press, UK, 2002). 

\bibitem{bray}A. J. Bray, \textit{Adv. Phys.} \textbf{51}, 481 (2002).

\bibitem{hansen}J.-P. Hansen and I.R. Mcdonald, 
\textit{Theory of Simple Liquids} (Academic Press, London, 2008).

\bibitem{allen}M.P. Allen and D.J. Tildsely, 
\textit{Computer Simulations of Liquids} (Clarendon, Oxford, 1987).

\bibitem{yeth1}K. Jagannathan and A. Yethiraj, \textit{Phys. Rev. Lett.} \textbf{93}, 015701 (2004).

\bibitem{yeth2}K. Jagannathan and A. Yethiraj, \textit{J. Chem. Phys.} \textbf{122}, 244506 (2005).

\bibitem{chen}A. Chen, E.H. Chimowitz, S. De and Y. Shapir, \textit{Phys. Rev. Lett.} 
\textbf{95}, 255701 (2005).

\bibitem{das2006}S.K. Das, M.E. Fisher, J.V. Sengers, J. Horbach, 
and K. Binder, \textit{Phys. Rev. Lett.} \textbf{97}, 025702 (2006).

\bibitem{dasjcp2006}S.K. Das, J. Horbach, K. Binder, M.E. Fischer and 
J.V. Sengers, \textit{J. Chem. Phys.} \textbf{125}, 024506 (2006).

\bibitem{roy1}S. Roy and S.K. Das, \textit{Europhys. Lett.} \textbf{94}, 
36001 (2011).

\bibitem{roy2}S. Roy and S.K. Das, \textit{J. Chem. Phys.}, \textbf{139},
 064505 (2013).

\bibitem{zinn}J. Zinn-Justin, \textit{Phys. Repts.} \textbf{344}, 159 (2001).

\bibitem{newrev} J.V. Sengers and R.A. Perkins, \textit{Fluids near Critical Points},
in \textit{Transport Properties of Fluids: Advances in Transport Properties}, eds. M.J. Assael, 
A.R.H. Goodwin, V. Vesovic and W.A. Wakeham (IUPAC, RSC Publishing, Cambridge, 2014), pp. 337-361.

\bibitem{hohenberg}P.C. Hohenberg and B.I. Halperin, 
\textit{Rev. Mod. Phys.} \textbf{49}, 435 (1977).

\bibitem{kadanoff}L.P. Kadanoff and J. Swift, \textit{Phys. Rev.} 
\textbf{166}, 89 (1968).

\bibitem{anisimov}M.A. Anisimov and J.V. Sengers, in 
\textit{Equations of State for Fluids and Fluid Mixtures}, 
ed. J.V. Sengers, R.F. Kayser, C.J. Peters and H.J. White, 
Jr. (Elsevier, Amsterdam, 2000) p.381.

\bibitem{onuki2}A. Onuki, \textit{Phys. Rev. E} \textbf{55}, 403 (1997).

\bibitem{olchowy}G.A. Olchowy and J.V. Sengers, \textit{Phys. Rev. Lett.} 
\textbf{61}, 15 (1988).

\bibitem{folk}R. Folk and G. Moser, \textit{Phys. Rev. Lett.} \textbf{75}, 
2706 (1995).

\bibitem{hao}H. Hao, R.A. Ferrell and J.K. Bhattacharjee, 
\textit{Phys. Rev. E.} \textbf{71}, 021201 (2005).

\bibitem{jkb1}J.K. Bhattacharjee, I. Iwanowski and
U. Kaatze, \textit{J. Chem. Phys.} \textbf{131}, 174502 (2009).

\bibitem{jkb2}J.K. Bhattacharjee, U. Kaatze and S.Z. Mirzaev, 
\textit{Rep. Progr. Phys.} \textbf{73}, 066601 (2010).

\bibitem{sengers}J.V. Sengers and J.G. Shanks, \textit{J.Stat. Phys.} 
\textbf{137}, 857 (2009).

\bibitem{burstyn1}H.C. Burstyn and J.V. Sengers, \textit{Phys. Rev. Lett.} 
\textbf{45}, 259 (1980).

\bibitem{burstyn2}H.C. Burstyn and J.V. Sengers, 
\textit{Phys. Rev. A} \textbf{25}, 448 (1982).

\bibitem{landau}D.P. Landau and K. Binder, \textit{A Guide to Monte 
Carlo Simulations in Statistical Physics}, 3rd Edition (Cambridge 
University Press, Cambridge, 2009). 

\bibitem{frenkel}D. Frenkel and B. Smit, \textit{Understanding 
Molecular Simulations: From Algorithm to Applications} 
(Academic Press, San Diego, 2002).

\bibitem{rapaport}D.C. Rapaport, \textit{The Art of Molecular Dynamics 
Simulations} (Cambridge University Press, Cambridge, UK, 2004).

\bibitem{stoya}S.D. Stayonav and R.D. Groot, \textit{J. Chem. Phys.} \textbf{122}, 114112.

\bibitem{niku}P. Nikunen, M. Karttunen and I. Vattulainen, \textit{Comp. Phys. Comm.} 
\textbf{153}, 407 (2003).

\bibitem{gomp}M. Ripoll, M. Mussawisade, R. G. Winkler and G. Gompper, 
\textit{Phys. Rev. E} \textbf{72}, 016701 (2005).

\bibitem{fisher_fss}M.E. Fisher, in \textit{Critical Phenomena}, 
edited by M.S. Green (Academic Press, London, 1971) p.1.

\bibitem{ferrell}R.A. Ferrell and J.K. Bhattacharjee, 
\textit{Phys. Rev. Lett.} \textbf{88}, 77 (1982).

\bibitem{jkbphysa}J.K. Bhattacharjee and R.A. Ferrell, \textit{Physica A} \textbf{250}, 83 (1998).

\bibitem{schmid}M.P. Allen and F. Schmid, \textit{Molecular Simulation} \textbf{33}, 
21 (2006).

\bibitem{dsf2007}S.K. Das, J.V. Sengers and M.E. Fisher, \textit{J. Chem. Phys.} \textbf{127},
144506 (2007).

\bibitem{mold1999} R.F. Berg, M.R. Moldover and G.A. Zimmerli, Phys. Rev. Lett. \textbf{82}, 920 (1999).


\end{thebibliography}
\end{document}